\input{aipcheck}

\documentclass[
    ,final            % use final for the camera ready runs
  ]
  {aipproc}

\layoutstyle{6x9}

\usepackage{amssymb}
\usepackage{amsmath}
\usepackage{braket}
\usepackage{doublespace}

\begin{document}

\title{Limit distribution for a time-inhomogeneous 2-state quantum walk}


\author{Takuya Machida${}^{\ast}$}{
  address={Research Fellow of Japan Society for the Promotion of Science\\
University of California, Berkeley, Department of Mathematics\\
Email: machida@stat.t.u-tokyo.ac.jp
}
}

\begin{abstract}
We consider 2-state quantum walks (QWs) on the line, which are defined by two matrices.
One of the matrices operates the walk in certain intervals.
In the usual QWs starting from the origin, localization does not occur at all.
However, our walk can be localized around the origin.
In this paper, we present some limit distributions for the walk.
\end{abstract}

\maketitle

%%%%%%%%%%%%%%%%%%%%%%%%%%%%%%%%%%%%%%%%%%%%
%% MAINMATTER
%%%%%%%%%%%%%%%%%%%%%%%%%%%%%%%%%%%%%%%%%%%%

\section{1.~~Introduction}
\label{intro}
The quantum walks (QWs) are considered as the quantum counterparts of a random walk.
The discrete-time QWs introduced by Aharonov et al. \cite{aharonov} and Meyer \cite{meyer} were intensively studied in Ambainis et al. \cite{ambainis_2001}.
On the other hand, the continuous-time models were defined by Farhi and Gutmann \cite{farhi}.
We concentrate on the discrete-time 2-state QWs in this work.
In relation with quantum computer, QWs are often investigated.
Since the quantum walker has ballistic behavior in probability distribution, quantum searches which are more fruitful than the classical ones can be designed by QWs \cite{childs_2009,childs_2002,reitzner,shenvi}.
In other research field, applications of the walks were also discussed \cite{abal_2008,mohseni_2008,oka}.
The asymptotic behaviors of the walks have been analyzed, and limit theorems have been obtained.
For example, Konno \cite{konno_2002_1,konno_2005_1} calculated the limit distribution of the usual 2-state walk (i.e., time-homogeneous model) on the line $\mathbb{Z}=\left\{\,0,\pm 1,\pm 2,\ldots\,\right\}$ which had a simple density function.
In the present paper, we consider a time-inhomogeneous 2-state quantum walk (QW) determined by two unitary matrices on the line. 
There are some results for time-inhomogeneous models \cite{banuls,machida_2010_1,romanelli_2009_1}.
Ba\~{n}uls et al. \cite{banuls} and Romanelli \cite{romanelli_2009_1} insisted localization for their models, respectively.
Machida and Konno \cite{machida_2010_1} gave convergence theorems to a 2-period QW with two unitary matrices and two special time-dependent walks.
We focus on calculation of the limit distribution which can be expressed by a combination of density functions. 
QWs whose limit distributions had similar structures to ours were treated in \cite{brun,miyazaki,segawa}.
Brun et al. \cite{brun} introduced and discussed the multi-state walks with the tensor-product representations.
A limit theorem for the multi-state walk with a special initial state was obtained in Segawa and Konno \cite{segawa}.
Miyazaki et al. \cite{miyazaki} also computed the limit distribution of the walk defined by the Wigner formula of rotation matrices.
If the number of states is more than four in these multi-state models, the limit distributions are depicted as a combination of density functions. 

This paper is organized as follows.
In Sect. 2, we define our walk.
Sect. 3 is devoted to calculation for the Fourier transform of amplitude.
We present limit distributions as our main result in Sect. 4.
A detail of the computation for the limit distributions is given in Appendix. 
By using the Fourier analysis introduced and studied by Grimmett et al. \cite{grimmett}, we obtain the limit distribution.
Summary is given in the final section.

%%%%%%%%%%%%%%%%%%%%%%%%%%%%%%%%%%%%%%%%%%%%%%%    DEFINITION     %%%%%%%%%%%%%%%%%%%%%%%%%%%%%%%%%%%%%%%%%%%%%%%%%%%%%%%%%%%

\section{2.~~Definition of our 2-state QWs}
\label{sec:2}
In this section we define our 2-state QWs on the line.
Let $\ket{x}$ ($x\in\mathbb{Z}$) be an infinite components vector which denotes the position of the walker.
Here,  $x$-th component of $\ket{x}$ is 1 and the other is 0.
Let $\ket{\psi_{t}(x)} \in \mathbb{C}^2$ be the amplitude of the walker at position $x$ at time $t\in\left\{0,1,2,\ldots\right\}$, where $\mathbb{C}$ is the set of complex numbers.
The walk at time $t$ is expressed by
\begin{equation}
 \ket{\Psi_t}=\sum_{x\in\mathbb{Z}}\ket{x}\otimes\ket{\psi_{t}(x)}.
\end{equation}
The time evolution of our walk is constructed by the following two unitary matrices:
\begin{equation}
  U=\left[\begin{array}{cc}
     \cos\theta &\sin\theta \\\sin\theta &-\cos\theta
	  \end{array}\right]
 =\left[\begin{array}{cc}
     c &s \\s &-c
	  \end{array}\right],\,
  H=\left[\begin{array}{cc}
     1&0\\ 0&-1
	  \end{array}\right],
\end{equation}
where $c=\cos\theta, s=\sin\theta$ and $\theta\in[0,2\pi)\,(\theta\neq 0,\frac{\pi}{2},\pi,\frac{3\pi}{2})$.
Moreover, we introduce four matrices:
\begin{equation}
 P=\left[\begin{array}{cc}
    c & s\\ 0&0
	 \end{array}\right],\,
 Q=\left[\begin{array}{cc}
    0&0\\ s &-c
	 \end{array}\right],\,
 P_1=\left[\begin{array}{cc}
	    1 & 0\\ 0&0
		 \end{array}\right],\,
 Q_1=\left[\begin{array}{cc}
	    0&0\\ 0 &-1
		 \end{array}\right].
\end{equation}
Then the evolution is determined as follows:\\
For $t=\tau+1,2(\tau+1),\ldots,m(\tau+1)$,
\begin{equation}
 \ket{\psi_{t}(x)}=P_1\ket{\psi_{t-1}(x+1)}+Q_1\ket{\psi_{t-1}(x-1)},\label{eq:te1}
\end{equation}
and for $t\neq \tau+1,2(\tau+1),\ldots,m(\tau+1)$,
\begin{equation}
 \ket{\psi_{t}(x)}=P\ket{\psi_{t-1}(x+1)}+Q\ket{\psi_{t-1}(x-1)},\label{eq:te2}
\end{equation}
with $m,\tau\in\left\{1,2,\ldots\right\}$.
Note that $P+Q=U$ and $P_1+Q_1=H$.
The probability that the quantum walker $X_t$ is at position $x$ at time $t$, $\mathbb{P}(X_t=x)$, is defined by
\begin{equation}
 \mathbb{P}(X_t=x)=\braket{\psi_t(x)|\psi_t(x)}.\label{eq:prob}
\end{equation}\label{eq:ft_amp}
The Fourier transform $\ket{\hat{\Psi}_{t}(k)}\,(k\in\left[-\pi,\pi\right))$ of $\ket{\psi_t(x)}$ is given by
\begin{equation}
 \ket{\hat{\Psi}_{t}(k)}=\sum_{x\in\mathbb{Z}} e^{-ikx}\ket{\psi_t(x)}.\label{eq:ft}
\end{equation}
By the inverse Fourier transform, we have
\begin{equation}
 \ket{\psi_t(x)}=\int_{-\pi}^{\pi}e^{ikx}\ket{\hat\Psi_{t}(k)}\,\frac{dk}{2\pi}.
\end{equation}
From Eqs. (\ref{eq:te1}), (\ref{eq:te2}) and (\ref{eq:ft}), the time evolution of $\ket{\hat{\Psi}_{t}(k)}$ becomes
\begin{equation}
 \ket{\hat{\Psi}_{t}(k)}=
  \left\{\begin{array}{ll}
   \hat H(k)\ket{\hat{\Psi}_{t-1}(k)}&(t=\tau+1,2(\tau+1),\ldots,m(\tau+1)),\\[2mm]
    \hat U(k)\ket{\hat{\Psi}_{t-1}(k)}&(t\neq\tau+1,2(\tau+1),\ldots,m(\tau+1)),
	 \end{array}\right.\label{eq:timeevo}
\end{equation}
where $\hat U(k)=R(k)U,\,\hat{H}(k)=R(k)H$ and
$
 R(k)=\left[\begin{array}{cc}
       e^{ik}&0\\
	     0&e^{-ik}
	    \end{array}\right]
$.
In order to analyze the walk after time $m(\tau+1)$, we focus on the probability distribution especially at time $(m+n)\tau+m \,(n\in\left\{1,2,\ldots\right\})$.
Then the Fourier transform of amplitude can be written as follows:
\begin{equation}
 \ket{\hat{\Psi}_{(m+n)\tau+m}(k)}=\hat U(k)^{n\tau} (\hat H(k)\hat U(k)^\tau)^m\ket{\hat\Psi_{0}(k)}.\label{eq:ft_amp}
\end{equation}
In Sect. 3, Eq. (\ref{eq:ft_amp}) will be expressed by the eigenvalues and eigenvectors of $\hat U(k)$.
In the present paper, we take the initial state as
\begin{equation}
 \ket{\psi_0(x)}=\left\{\begin{array}{ll}
		 \!{}^T[\,\alpha, \,\beta\,]& (x=0),\\[2mm]
			\!{}^T[\,0,\,0\,]& (x\neq 0),
		       \end{array}\right.\label{eq:ini}
\end{equation}
where $|\alpha|^2+|\beta|^2=1$ and $T$ is the transposed operator.
We should note that $\ket{\hat\Psi_{0}(k)}=\ket{\psi_0(0)}$.

%%%%%%%%%%%%%%%%%%%%%%%%%%%%%%%%%%%%%%%%   AMPLITUDE  %%%%%%%%%%%%%%%%%%%%%%%%%%%%%%%%%%%%%%%%%%%%
\section{3.~~The Fourier transform of amplitude expressed by the eigenvalues and eigenvectors of $\hat U(k)$}
\label{sec:3}

In this section we rewrite $\ket{\hat\Psi_{(m+n)\tau+m}(x)}$ in Eq. (\ref{eq:ft_amp}) by using the eigenvalues and eigenvectors of $\hat U(k)$ whose formulae will be given in Appendix. 
At first, the Fourier transform at time $m(\tau+1)$ can be described as follows:
\begin{equation}
  \ket{\hat\Psi_{m(\tau+1)}(k)}=\sum_{j=1}^2 \braket{v_j(k)|\hat\Psi_{m(\tau+1)}(k)}\ket{v_j(k)}=\sum_{j=1}^2 a_{j,m}(k)\ket{v_j(k)},\label{eq:ft_amplitude}
\end{equation}
where $a_{j,m}(k)=\braket{v_j(k)|\hat\Psi_{m(\tau+1)}(k)}$ and $\ket{v_j(k)}(j\in\left\{1,2\right\})$ are the normalized eigenvectors corresponding to the eigenvalues $\lambda_j(k)$ of the unitary matrix $\hat U(k)$.
In particular, since the initial state $\ket{\psi_0(x)}$ is given by Eq. (\ref{eq:ini}) in our model, we have $a_{j,0}(k)=\braket{v_j(k)|\hat\Psi_0(k)}=\braket{v_j(k)|\psi_0(0)}$.
From Eq. (\ref{eq:ft_amplitude}), we see that
\begin{align}
 \ket{\hat\Psi_{m(\tau+1)}(k)}=&\hat H(k)\hat U(k)^\tau\ket{\hat\Psi_{(m-1)(\tau+1)}(k)}\nonumber\\
 =&\sum_{j_1=1}^2a_{j_1,m-1}(k)\lambda_{j_1}(k)^\tau \hat H(k)\ket{v_{j_1}(k)}\nonumber\\
 =&\sum_{j_1=1}^2\sum_{j_2=1}^2 a_{j_1,m-1}(k)\lambda_{j_1}(k)^\tau A_{j_2j_1}(k)\ket{v_{j_2}(k)},
\end{align}
where $A_{j_2j_1}(k)=\braket{v_{j_2}(k)|\hat{H}(k)|v_{j_1}(k)}\, (j_1,j_2\in\left\{1,2\right\})$.
Therefore we get
\begin{align}
 \left[\begin{array}{c}
  a_{1,m}(k)\\a_{2,m}(k)
       \end{array}\right]
 =&\left[\begin{array}{cc}
   \lambda_1(k)^\tau A_{11}(k)&\lambda_2(k)^\tau A_{12}(k) \\
	 \lambda_1(k)^\tau A_{21}(k)&\lambda_2(k)^\tau A_{22}(k)
	\end{array}\right]
 \left[\begin{array}{c}
  a_{1,m-1}(k)\\ a_{2,m-1}(k)
       \end{array}\right]\nonumber\\
 =&\left[\begin{array}{cc}
   \lambda_1(k)^\tau A_{11}(k)&\lambda_2(k)^\tau A_{12}(k) \\
	 \lambda_1(k)^\tau A_{21}(k)&\lambda_2(k)^\tau A_{22}(k)
	\end{array}\right]^m
 \left[\begin{array}{c}
  a_{1,0}(k)\\ a_{2,0}(k)
       \end{array}\right].
\end{align}
Moreover, the coefficients $a_{j,m}(k)$ of the eigenvectors $\ket{v_j(k)}$ can be computed as follows:
\begin{align}
 a_{1,m}(k)=&\frac{1}{z_2(k)-z_1(k)}\left\{(-1)^\tau(z_2(k)^{m-1}-z_1(k)^{m-1})a_{1,0}(k)\right.\nonumber\\
 &\hspace{2.5cm}+\lambda_1(k)^\tau A_{11}(k)(z_2(k)^m-z_1(k)^m)a_{1,0}(k)\nonumber\\
 &\hspace{2.5cm}\left.+\lambda_2(k)^\tau A_{12}(k)(z_2(k)^m-z_1(k)^m)a_{2,0}(k)\right\},\label{eq:a_{1,m}}\\
 a_{2,m}(k)=&\frac{1}{z_2(k)-z_1(k)}\left\{\lambda_1(k)^\tau A_{12}(k)(z_2(k)^m-z_1(k)^m)a_{1,0}(k)\right.\nonumber\\
 &\hspace{2.5cm}+(z_2(k)^{m+1}-z_1(k)^{m+1})a_{2,0}(k)\nonumber\\
 &\hspace{2.5cm}\left.-\lambda_1(k)^\tau A_{11}(k)(z_2(k)^m-z_1(k)^m)a_{2,0}(k)\right\},\label{eq:a_{2,m}}
\end{align}
where $z_j(k)\,(j\in\left\{1,2\right\})$ are the eigenvalues of the following matrix:
\begin{equation}
 \left[\begin{array}{cc}
   \lambda_1(k)^\tau A_{11}(k)&\lambda_2(k)^\tau A_{12}(k) \\
	 \lambda_1(k)^\tau A_{21}(k)&\lambda_2(k)^\tau A_{22}(k)
	\end{array}\right],
\end{equation}
that is,
\begin{equation}
 z_j(k)=\nu(k)-(-1)^j\sqrt{\nu(k)^2+(-1)^\tau},
\end{equation}
and
\begin{equation}
 \nu(k)=\frac{\lambda_1(k)^\tau A_{11}(k)-(-1)^\tau\overline{\lambda_1(k)^\tau A_{11}(k)}}{2}.
\end{equation}
Hence, for $a_{m,j}(k)$ given by Eqs. (\ref{eq:a_{1,m}}) and (\ref{eq:a_{2,m}}), Eq. (\ref{eq:ft_amp}) becomes
\begin{equation}
 \ket{\hat\Psi_{(m+n)\tau+m}(k)}=(\hat U(k)^\tau)^n\ket{\hat\Psi_{m(\tau+1)}(k)}=\sum_{j=1}^2 a_{j,m}(k)\lambda_j(k)^{n\tau}\ket{v_j(k)}.\label{eq:psi_(n+2)t+2}
\end{equation}

%\clearpage

%%%%%%%%%%%%%%%%%%%%%%%%%%%%%%%%%%%%%%%%%%%%%%%%%    LIMIT DISTRIBUTION     %%%%%%%%%%%%%%%%%%%%%%%%%%%%%%%%%%%%%%%%%%%%%%%%%%%%%%%%%%%%%

\section{4.~~Limit distributions for the walk}
\label{sec:4}

\begin{figure}[h]
% \begin{center}
 \begin{minipage}{50mm}
  \begin{center}
   \includegraphics[scale=0.3]{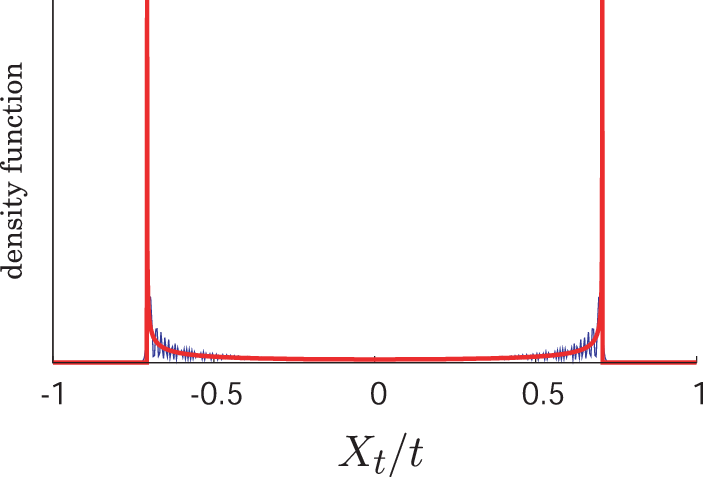}\\
   {(a) $\ket{\psi_0(0)}={}^T[1/\sqrt{2}\,,i/\sqrt{2}\,]$}
  \end{center}
 \end{minipage}%\hspace{1cm}
 \begin{minipage}{50mm}
  \begin{center}
   \includegraphics[scale=0.3]{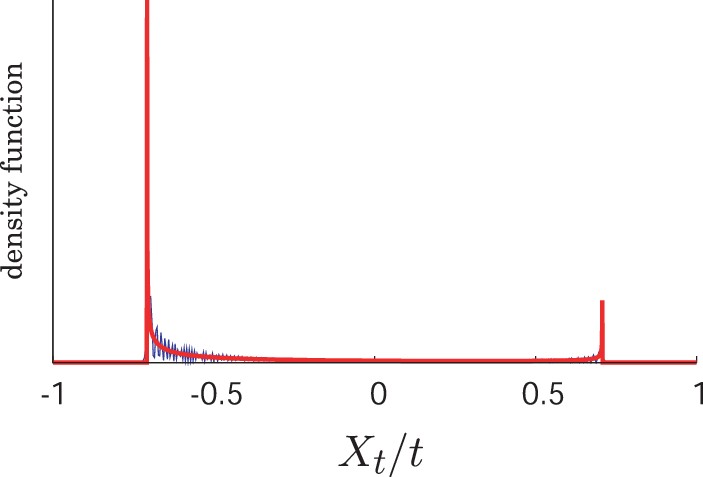}\\
   {(b) $\ket{\psi_0(0)}={}^T[1,0]$}
  \end{center}
 \end{minipage}
 \caption{Comparison between the limit density functions (thick line) and the probability distributions at time $t=500$ as $\tau=500$ (thin line) with $\theta=\pi/4$.}
  \label{fig:dis_usual}
% \end{center}
\end{figure}

In this section the limit distributions for some cases, one of which will be computed in detail in Appendix, are presented.
Firstly, we show the limit distribution for the usual walks which was given by Konno \cite{konno_2002_1,konno_2005_1}.
Since the matrix $H$ does not operate our walk till time $\tau$, we can obtain the convergence in distribution of the walk $X_\tau/\tau$ as the limit theorem for the usual 2-state walks.
That is, for the walk, we have
\begin{equation}
 \lim_{\tau\to\infty}\mathbb{P}\left(\frac{X_\tau}{\tau}\leq x\right)=\int_{-\infty}^x f_K(y;c)\left[1-\left\{|\alpha|^2-|\beta|^2+\frac{(\alpha\overline{\beta}+\overline{\alpha}\beta)s}{c}\right\}y\right]\,dy,
\end{equation}
where
\begin{equation}
 f_K(x;c)=\frac{|s|}{\pi(1-x^2)\sqrt{c^2-x^2}}I_{(-|c|,|c|)}(x),
\end{equation}
and $I_A(x)=1$ if $x\in A$, $I_A(x)=0$ if $x\notin A$.
Fig.\ref{fig:dis_usual} depicts the comparison between the limit density functions and the probability distributions at time 500 for usual walks with $\theta=\pi/4$.

%\clearpage

\subsection{$m=1$ case}
\label{subsec:4-1}
From now, we propose our results for the walks.
At first, we concentrate on $m=1$ case and consider the probability distribution at time $t=(n+1)\tau+1$.
The limit density functions as $\tau\to\infty$ are calculated for $n=1$ case in which the limit distribution was computed by Machida \cite{machida_2010_3} and $n=2,3,\ldots$ case, respectively.

\begin{figure}[h]
% \begin{center}
 \begin{minipage}{50mm}
  \begin{center}
   \includegraphics[scale=0.3]{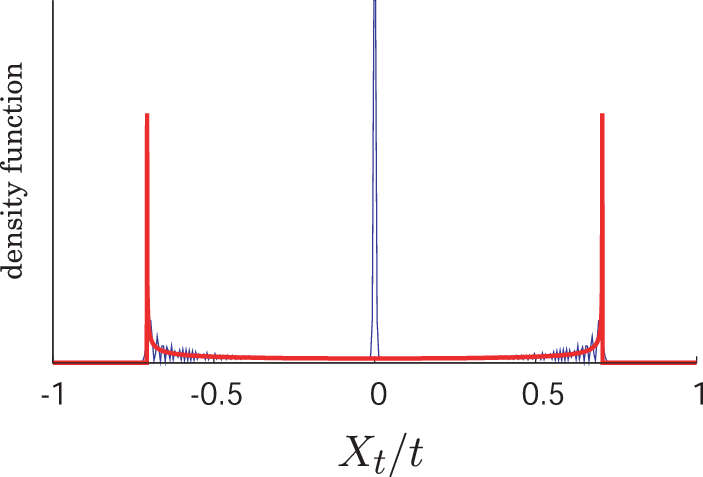}\\
   {(a) $\ket{\psi_0(0)}={}^T[1/\sqrt{2}\,,i/\sqrt{2}\,]$}
  \end{center}
 \end{minipage}%\hspace{1cm}
 \begin{minipage}{50mm}
  \begin{center}
   \includegraphics[scale=0.3]{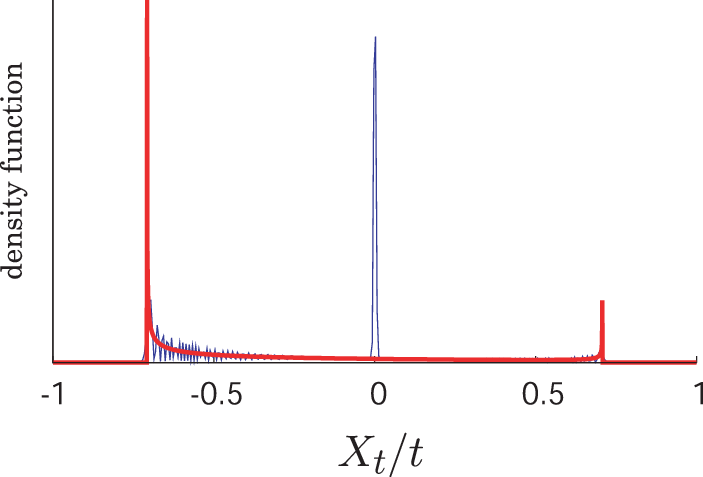}\\
   {(b) $\ket{\psi_0(0)}={}^T[1,0]$}
  \end{center}
 \end{minipage}
 \caption{Comparison between the limit density functions (thick line) and the probability distributions at time $t=401$ as $\tau=200$ (thin line) with $\theta=\pi/4,m=1,n=1$.}
  \label{fig:dis_1-1}
% \end{center}
\end{figure}

\noindent\underline{$n=1$ case}\\
For $n=1$, Machida \cite{machida_2010_3} obtained the limit distribution expressed by both a $\delta$-function and a density function as follows:
\begin{equation}
 \lim_{\tau\to\infty}\mathbb{P}\left(\frac{X_{2\tau+1}}{2\tau}\leq x\right)=\int^{x}_{-\infty}\Delta_1\delta_0(y)+f_K(y;c)M_1(y)\,dy,
\end{equation}
where $\delta_0(x)$ is Dirac's $\delta$-function at the origin and
\begin{align}
 \Delta_1=&\frac{s^2}{1+|s|},\label{eq:delta_1-1}\\
 M_1(x)=&\left[1-\left\{|\alpha|^2-|\beta|^2+\frac{(\alpha\overline{\beta}+\overline{\alpha}\beta)s}{c}\right\}x\right]\left(1-\frac{s^2}{c^2}x^2\right).
\end{align}
We should note that $\Delta_1$ doesn't depend on the initial state $\ket{\psi_0(0)}={}^T\left[\alpha,\beta\right]$.
Assuming the initial state as $\ket{\psi_0(0)}={}^T[\frac{1}{\sqrt{2}}\,,\frac{i}{\sqrt{2}}\,]$ and $\ket{\psi_0(0)}={}^T[1,0]$, we get the limit densities and the probability distributions as shown by Fig. \ref{fig:dis_1-1}.
%\vspace{-4mm}
\clearpage

%\clearpage
%\vspace{-3mm}

\begin{figure}[h]
% \begin{center}
 \begin{minipage}{50mm}
  \begin{center}
   \includegraphics[scale=0.3]{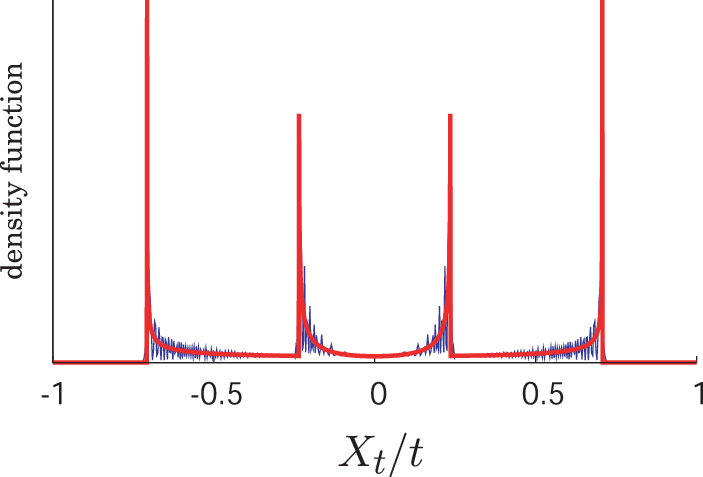}\\
   {(a) $\ket{\psi_0(0)}={}^T[1/\sqrt{2}\,,i/\sqrt{2}\,]$}
  \end{center}
 \end{minipage}%\hspace{1cm}
 \begin{minipage}{50mm}
  \begin{center}
   \includegraphics[scale=0.3]{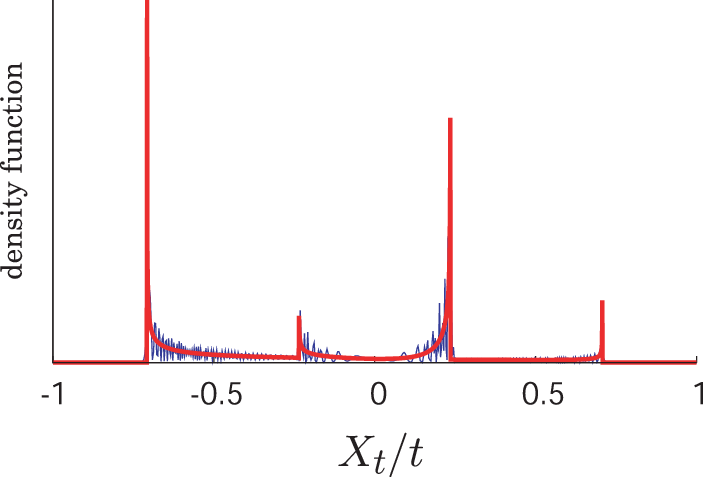}\\
   {(b) $\ket{\psi_0(0)}={}^T[1,0]$}
  \end{center}
 \end{minipage}
 \caption{Comparison between the limit density functions (thick line) and the probability distributions at time $t=601$ as $\tau=200$ (thin line) with $\theta=\pi/4,m=1,n=2$.}
  \label{fig:dis_1-2}
% \end{center}
\end{figure}

\noindent\underline{$n=2,3,\ldots$ case}\\
For $n=2,3,\ldots$, the limit distribution is denoted by two density functions as follows:
\begin{equation}
 \lim_{\tau\to\infty}\mathbb{P}\left(\frac{X_{(n+1)\tau+1}}{(n+1)\tau}\leq x\right)=\int^{x}_{-\infty}f_K(y;c)M_1(y)+f_K(\frac{n+1}{n-1}y;c)B_{1}(\frac{n+1}{n-1}y;n)\,dy,
\end{equation}
where
\begin{align}
 B_{1}(x;n)=&\frac{n+1}{n-1}\left[1+\left\{|\alpha|^2-|\beta|^2+\frac{(\alpha\overline{\beta}+\overline{\alpha}\beta)s}{c}\right\}x\right]\frac{s^2}{c^2}x^2.
\end{align}
We show the symmetric and asymmetric distributions for $n=2$ in Fig. \ref{fig:dis_1-2}.
%\vspace{-4mm}

\subsection{$m=2$ case}
\label{subsec:4-2}
Next we consider $m=2$ case and focus on the probability distribution at time $t=(n+2)\tau+2$.
In this case we compute the limit density functions as $\tau\to\infty$ for $n=1$, $n=2$ and $n=3,4,\ldots$ cases, respectively.
For $n=3,4,\ldots$, the calculation will be written in Appendix.

\begin{figure}[h]
% \begin{center}
 \begin{minipage}{50mm}
  \begin{center}
   \includegraphics[scale=0.3]{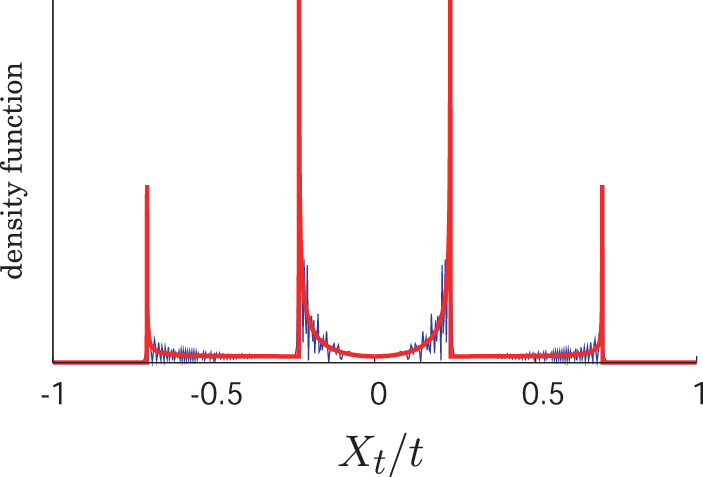}\\
   {(a) $\ket{\psi_0(0)}={}^T[1/\sqrt{2}\,,i/\sqrt{2}\,]$}
  \end{center}
 \end{minipage}%\hspace{1cm}
 \begin{minipage}{50mm}
  \begin{center}
   \includegraphics[scale=0.3]{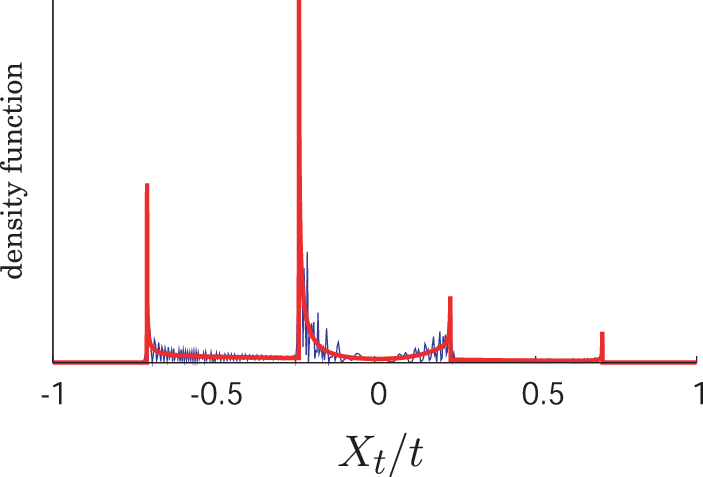}\\
   {(b) $\ket{\psi_0(0)}={}^T[1,0]$}
  \end{center}
 \end{minipage}
 \caption{Comparison between the limit density functions (thick line) and the probability distributions at time $t=602$ as $\tau=200$ (thin line) with $\theta=\pi/4,m=2,n=1$.}
  \label{fig:dis_2-1}
% \end{center}
\end{figure}

\noindent\underline{$n=1$ case}\\
For $n=1$, the limit distribution has a combination of two density functions as follows:
\begin{equation}
 \lim_{\tau\to\infty}\mathbb{P}\left(\frac{X_{3\tau+2}}{3\tau}\leq x\right)=\int^{x}_{-\infty}f_K(y;c)M_2(y)+f_K(3y;c)M_3(3y)\,dy,
\end{equation}
where
\begin{align}
 M_2(x)=&\left[1-\left\{|\alpha|^2-|\beta|^2+\frac{(\alpha\overline{\beta}+\overline{\alpha}\beta)s}{c}\right\}x\right]\left(1-\frac{s^2}{c^2}x^2\right)^2,\\
 M_3(x)=&\frac{3s^2}{c^2}x^2\left[2-\frac{2(\alpha\overline{\beta}+\overline{\alpha}\beta)s(1+s^2)}{c^3}x-\frac{s^2}{c^2}x^2\right.\nonumber\\
 &\left.-3\left\{|\alpha|^2-|\beta|^2+\frac{(\alpha\overline{\beta}+\overline{\alpha}\beta)s}{c}\right\}\frac{s^2}{c^2}x^3
 +\frac{4(\alpha\overline{\beta}+\overline{\alpha}\beta)s^3}{c^3}\frac{x}{1-x^2}\right].
\end{align}
When we take the initial state as $\ket{\psi_0(0)}={}^T[\frac{1}{\sqrt{2}}\,,\frac{i}{\sqrt{2}}\,]$ and $\ket{\psi_0(0)}={}^T[1,0]$, the distributions are shown in Fig. \ref{fig:dis_2-1}

%\clearpage

\begin{figure}[h]
% \begin{center}
 \begin{minipage}{50mm}
  \begin{center}
   \includegraphics[scale=0.3]{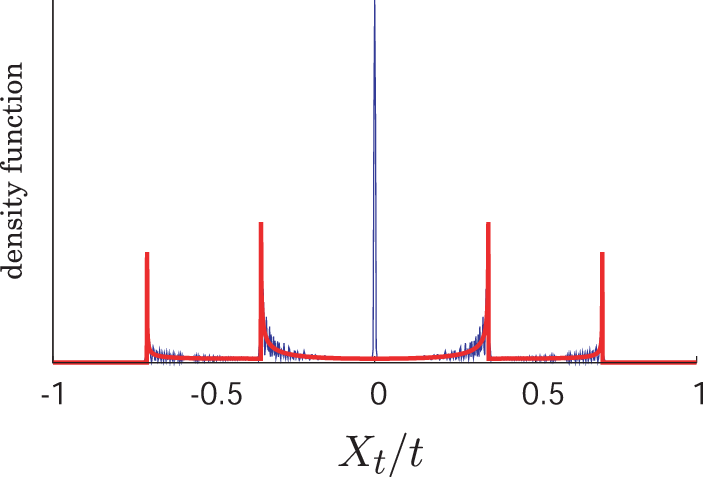}\\
   {(a) $\ket{\psi_0(0)}={}^T[1/\sqrt{2}\,,i/\sqrt{2}\,]$}
  \end{center}
 \end{minipage}%\hspace{1cm}
 \begin{minipage}{50mm}
  \begin{center}
   \includegraphics[scale=0.3]{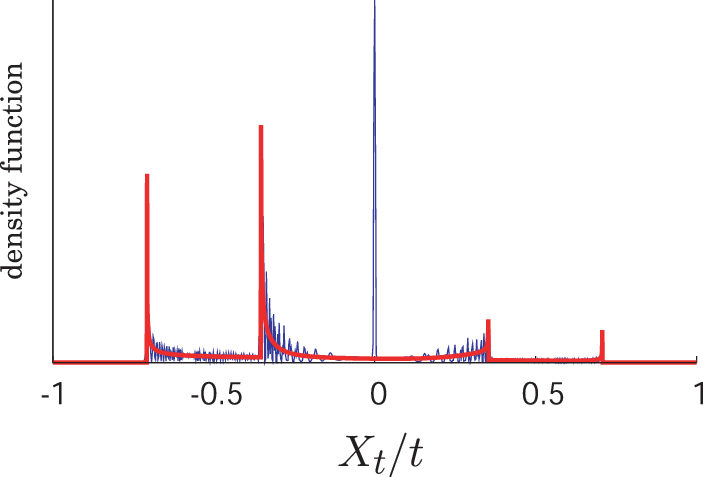}\\
   {(b) $\ket{\psi_0(0)}={}^T[1,0]$}
  \end{center}
 \end{minipage}
 \caption{Comparison between the limit density functions (thick line) and the probability distributions at time $t=802$ as $\tau=200$ (thin line) with $\theta=\pi/4,m=2,n=2$.}
  \label{fig:dis_2-2}
% \end{center}
\end{figure}

\noindent\underline{$n=2$ case}\\
For $n=2$, we can describe the limit distribution with a $\delta$-function and two density functions as follows:
\begin{equation}
 \lim_{\tau\to\infty}\mathbb{P}\left(\frac{X_{4\tau+2}}{4\tau}\leq x\right)=\int^{x}_{-\infty}\Delta_2\delta_0(y)+f_K(y;c)M_2(y)+f_K(2y;c)B_2(2y;2)\,dy,
\end{equation}
where
\begin{align}
 \Delta_2=&\frac{c^2|s|}{2}+\frac{(s^2-c^2)|s|(1-|s|)^2}{2c^4},\label{eq:delta_2-2}\\
 B_2(x;n)=&\frac{n+2}{n}\frac{s^2}{c^2}x^2\left[1+\left\{|\alpha|^2-|\beta|^2-\frac{(\alpha\overline{\beta}+\overline{\alpha}\beta)s(3s^2+1)}{c^3}\right\}x\right.\nonumber\\
 &\left.-4\left\{|\alpha|^2-|\beta|^2+\frac{(\alpha\overline{\beta}+\overline{\alpha}\beta)s}{c}\right\}\frac{s^2}{c^2}x^3+\frac{4(\alpha\overline{\beta}+\overline{\alpha}\beta)s^3}{c^3}\frac{x}{1-x^2}\right].
\end{align}
We should remark that $\Delta_2$ is independent from the initial state $\ket{\psi_0(0)}={}^T\left[\alpha,\beta\right]$.
Fig. \ref{fig:dis_2-2} presents the limit density functions and the probability distributions.

%\clearpage

\begin{figure}[h]
% \begin{center}
 \begin{minipage}{50mm}
  \begin{center}
   \includegraphics[scale=0.3]{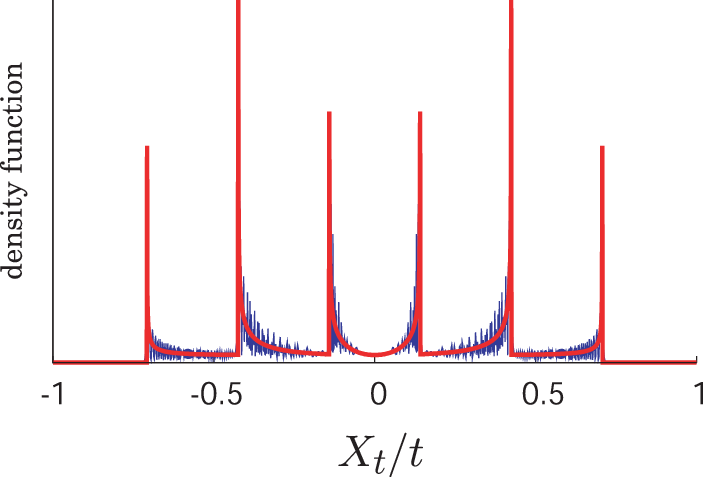}\\
   {(a) $\ket{\psi_0(0)}={}^T[1/\sqrt{2}\,,i/\sqrt{2}\,]$}
  \end{center}
 \end{minipage}%\hspace{1cm}
 \begin{minipage}{50mm}
  \begin{center}
   \includegraphics[scale=0.3]{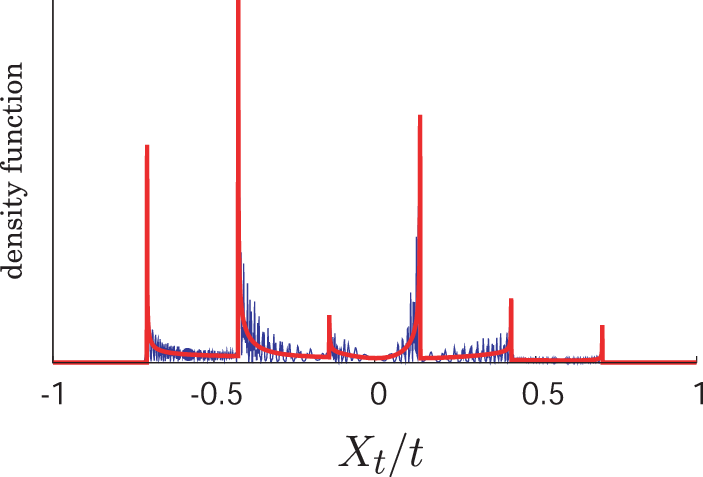}\\
   {(b) $\ket{\psi_0(0)}={}^T[1,0]$}
  \end{center}
 \end{minipage}
 \caption{Comparison between the limit density functions (thick line) and the probability distributions at time $t=1002$ as $\tau=200$ (thin line) with $\theta=\pi/4,m=2,n=3$.}
  \label{fig:dis_2-3}
% \end{center}
\end{figure}

\noindent\underline{$n=3,4,\ldots$ case}\\
For $n=3,4,\ldots$, the limit distribution is expressed by three density functions as follows:
\begin{align}
 &\lim_{\tau\to\infty}\mathbb{P}\left(\frac{X_{(n+2)\tau+2}}{(n+2)\tau}\leq x\right)\nonumber\\
=&\int^{x}_{-\infty}f_K(y;c)M_{2}(y)+f_K(\frac{n+2}{n}y;c)B_2(\frac{n+2}{n}y;n)+f_K(\frac{n+2}{n-2}y;c)B_3(\frac{n+2}{n-2}y;n)\,dy,
\end{align}
where
\begin{align}
 B_3(x;n)=&\frac{n+2}{n-2}\left[1+\left\{|\alpha|^2-|\beta|^2+\frac{(\alpha\overline{\beta}+\overline{\alpha}\beta)s}{c}\right\}x\right]\frac{s^2}{c^2}x^2\left(1-\frac{s^2}{c^2}x^2\right).
\end{align}
We draw the comparison  between the limit density functions and the probability distributions for $n=3$ in Fig. \ref{fig:dis_2-3}.
%\clearpage

\begin{figure}[h]
% \begin{center}
 \begin{minipage}{50mm}
  \begin{center}
   \includegraphics[scale=0.3]{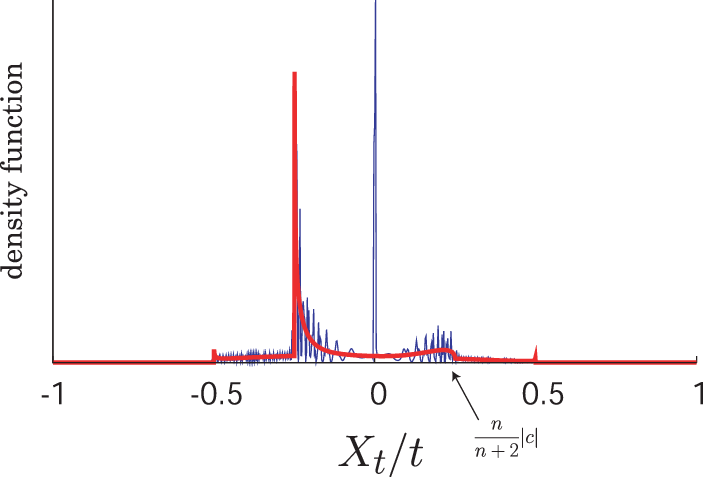}\\
   {(a) time $t=802$\\($\tau=200,m=2,n=2$)}
  \end{center}
 \end{minipage}%\hspace{1cm}
 \begin{minipage}{50mm}
  \begin{center}
   \includegraphics[scale=0.3]{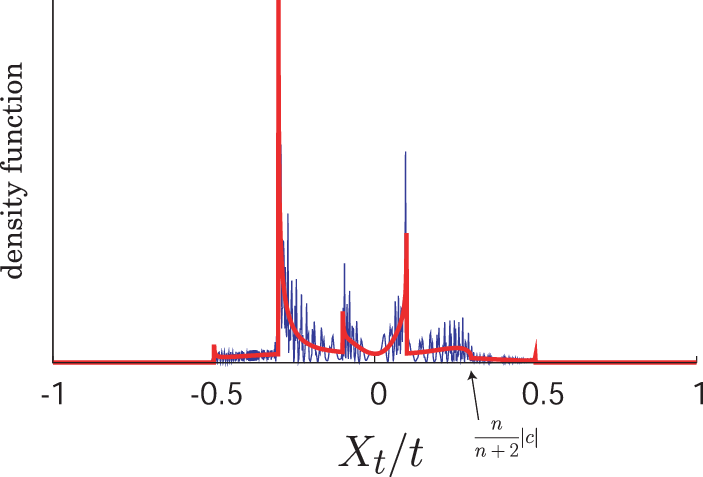}\\
   {(b) time $t=1002$\\($\tau=200,m=2,n=3$)}
  \end{center}
 \end{minipage}
 \caption{Comparison between the limit density functions (thick line) and the probability distributions (thin line) as the initial state $\ket{\psi_0(0)}={}^T[1,0]$ with $\theta=\pi/3$.}
  \label{fig:dis_interesting}
% \end{center}
\end{figure}

\noindent We should note that the function $f_K(\frac{n+2}{n}x;c)B_2(\frac{n+2}{n}x;n)$ can be finite at either $x=\frac{n}{n+2}|c|$ or $x=-\frac{n}{n+2}|c|$.
For example, if $\theta=\frac{\pi}{3}$ and $\ket{\psi_0(0)}={}^T[1,0]$, we see that
\begin{equation}
 \lim_{x\to\frac{n}{n+2}|c|}f_K(\frac{n+2}{n}x;c)B_2(\frac{n+2}{n}x;n)=0.\label{eq:fb_finite}
\end{equation}
Fig. \ref{fig:dis_interesting} shows typical examples of the distributions in the case when Eq. (\ref{eq:fb_finite}) is realized.

%%%%%%%%%%%%%%%%%%%%%%%%%        SUMMARY       %%%%%%%%%%%%%%%%%%%%%%%%%%%%%%
\section{5.~~Summary}
\label{sec:5}

In the final section, we conclude and discuss the probability distributions of our walks.
For the usual 2-state walk defined by $U$, the limit probability distribution of $X_\tau/\tau$ as $\tau\to\infty$ has a simple density function.
On the other hand, if another matrix $H$ operates the walk in certain intervals, the limit density function can be described by a $\delta$-function and a combination of density functions.
We found out a 2-state model whose limit distributions had the similar structures to the multi-state walks with more than four states defined in \cite{brun,miyazaki,segawa}.
In Sect. 4, we presented the limit densities for some cases.
Both coefficients $\Delta_1$ and $\Delta_2$ of the $\delta$-function given by Eqs. (\ref{eq:delta_1-1}) and (\ref{eq:delta_2-2}) respectively were particularly independent from the initial state $\ket{\psi_0(0)}={}^T[\alpha,\beta]$.
One of the further problems for our model is calculation of the limit distribution for general $m,n \,(m=3,4,\ldots)$ case and $m,n\to\infty$ case.

%%%%%%%%%%%%%%%%%%%%%%%%%%%%%%%%%%%%%%%%%%%%     APPENDIX     %%%%%%%%%%%%%%%%%%%%%%%%%%%%%%%%%%%%%%%%%%%%%%%%%%%%%
\appendix
\section{Appendix}
\section{A.~~Calculation of the limit distribution}
\label{sec:app}

In this appendix we calculate the limit distribution for $m=2,n=3,4,\ldots$ case.
For the other cases, the distributions can be obtained similarly.
Our approach is based on the Fourier analysis applied to QWs by Grimmett et al. \cite{grimmett}.
For $t=(n+2)\tau+2$, we concentrate on the characteristic function $E(e^{izX_t/(n+2)\tau})$ as $\tau\rightarrow\infty$, where $E(X)$ denotes the expected value of $X$.
At first, the eigenvalues $\lambda_j(k)\,(j\in\left\{1,2\right\})$ of $\hat U(k)$ are computed as
\begin{equation}
 \lambda_j(k)=-(-1)^j\sqrt{1-c^2\sin^2 k}+ic\sin k.
\end{equation}
The normalized eigenvector $\ket{v_j(k)}$ corresponding to $\lambda_j(k)$ is
\begin{equation}
  \ket{v_j(k)}=\sqrt{\frac{\sqrt{1-c^2\sin^2 k}-(-1)^j c\cos k}{2s^2\sqrt{1-c^2\sin^2 k}}}
   \left[\begin{array}{c}
    se^{ik}\\ -(-1)^j\sqrt{1-c^2\sin^2 k}-c\cos k
	 \end{array}\right].
\end{equation}
By using the eigenvalues and eigenvectors, Eq. (\ref{eq:psi_(n+2)t+2}) is written as
\begin{align}
 \ket{\hat\Psi_t(k)}=&(\hat U(k)^\tau)^n(\hat H(k)\hat U(k)^\tau)^2\ket{\hat\Psi_{0}(k)}\nonumber\\
 =&\Bigl\{\lambda_1(k)^{(n+2)\tau}a_{1,0}A_{11}^2-\lambda_1(k)^{(n-2)\tau}a_{2,0}\overline{A_{11}}A_{12}\nonumber\\
 &+(-1)^\tau\lambda_1(k)^{n\tau}(a_{1,0}A_{12}+a_{2,0}A_{11})A_{12}\Bigr\}\ket{v_1(k)}\nonumber\\
 &+\Bigl\{\lambda_2(k)^{(n+2)\tau}a_{2,0}\overline{A_{11}}^2+\lambda_2(k)^{(n-2)\tau}a_{1,0}A_{11}A_{12}\nonumber\\
 &+(-1)^\tau\lambda_2(k)^{n\tau}(a_{2,0}A_{12}-a_{1,0}\overline{A_{11}})A_{12}\Bigr\}\ket{v_2(k)},
\end{align}
where $a_{j,0}=a_{j,0}(k)=\braket{v_j(k)|\hat\Psi_0(k)}$ and $A_{j_1j_2}=A_{j_1j_2}(k)=\braket{v_{j_1}(k)|\hat H(k)|v_{j_2}(k)}\,(j_1,j_2\in\left\{1,2\right\})$.
Remark that $\ket{\hat\Psi_0(k)}=\ket{\psi_0(0)}={}^T[\alpha,\beta]$ (see Eq. (\ref{eq:ini})).
The {\it r}-th moment of $X_t$ is expressed as follows:
\begin{align}
 &E(X_t^r)=\sum_{x\in \mathbb{Z}}x^r \mathbb{P}(X_t=x)=\int_{-\pi}^{\pi}\bra{\hat\Psi_t(k)}\left(D^r\ket{\hat\Psi_t(k)}\right)\,\frac{dk}{2\pi}\nonumber\\
 =&((n+2)\tau)_r\int_{-\pi}^{\pi}\left\{|a_{1,0}A_{11}^2|^2 h_1(k)^r+|a_{2,0}\overline{A_{11}^2}|^2 h_2(k)^r\right\}\frac{dk}{2\pi}\nonumber\\
 &+((n-2)\tau)_r\int_{-\pi}^{\pi} \Bigl\{|a_{2,0}\overline{A_{11}}A_{12}|^2 h_1(k)^r+|a_{1,0}A_{11}A_{12}|^2 h_2(k)^r\Bigr\}\frac{dk}{2\pi}\nonumber\\
 &+(n\tau)_r \int_{-\pi}^{\pi} \Bigl\{|(a_{1,0}A_{12}+a_{2,0}A_{11})A_{12}|^2 h_1(k)^r\nonumber\\
 &\hspace{2cm}+|\left(a_{2,0}A_{12}-a_{1,0}\overline{A_{11}}\right)A_{12}|^2 h_2(k)^r\Bigr\}\frac{dk}{2\pi}\,+o(\tau^r),
\end{align}
with $h_j(k)=D\lambda_j(k)/\lambda_j(k)$, $D=i(d/dk)$ and $(t)_r=t(t-1)\times\cdots\times(t-r+1)$.
Noting that
\begin{align}
 h_{j}(k)=&(-1)^j\frac{c\cos k}{\sqrt{1-c^2\sin^2 k}},\\
 A_{11}(k)=&-\overline{A_{22}(k)}=\frac{c\cos^2 k+i\sin k\sqrt{1-c^2\sin^2 k}}{\sqrt{1-c^2\sin^2 k}},\\
 A_{12}(k)=&A_{21}(k)=\frac{|s|\cos k}{\sqrt{1-c^2\sin^2 k}},
\end{align}
in a similar way to Machida \cite{machida_2010_3}, we see that
\begin{align}
 \lim_{\tau\to\infty}E\left[\left(\frac{X_t}{(n+2)\tau}\right)^r\right]
 =&\int_{-\pi}^{\pi}\left\{|a_{1,0}A_{11}^2|^2 h_1(k)^r+|a_{2,0}\overline{A_{11}^2}|^2 h_2(k)^r\right\}\frac{dk}{2\pi}\nonumber\\
 &+\int_{-\pi}^{\pi} \left\{|a_{2,0}\overline{A_{11}}A_{12}|^2 \left(\frac{n-2}{n+2}h_1(k)\right)^r\right.\nonumber\\
 &\hspace{1cm}+\left.|a_{1,0}A_{11}A_{12}|^2 \left(\frac{n-2}{n+2}h_2(k)\right)^r\right\}\frac{dk}{2\pi}\nonumber\\
 &+\int_{-\pi}^{\pi} \left\{|(a_{1,0}A_{12}+a_{2,0}A_{11})A_{12}|^2 \left(\frac{n}{n+2}h_1(k)\right)^r\right.\nonumber\\
 &+\left.|\left(a_{2,0}A_{12}-a_{1,0}\overline{A_{11}}\right)A_{12}|^2 \left(\frac{n}{n+2}h_2(k)\right)^r\right\}\frac{dk}{2\pi}\nonumber\\
 =&\int_{-\infty}^{\infty} x^r \Biggl\{ f_K(x;c)M_{2}(x)+f_K(\frac{n+2}{n}x;c)B_2(\frac{n+2}{n}x;n)\nonumber\\
 &\hspace{1.5cm}+f_K(\frac{n+2}{n-2}x;c)B_3(\frac{n+2}{n-2}x;n)\Biggr\}\,dx,\label{eq:r-th_mom}
\end{align}
where
\begin{align}
 f_K(x;c)=&\frac{|s|}{\pi(1-x^2)\sqrt{c^2-x^2}}I_{(-|c|,|c|)}(x),\\
 M_2(x)=&\left[1-\left\{|\alpha|^2-|\beta|^2+\frac{(\alpha\overline{\beta}+\overline{\alpha}\beta)s}{c}\right\}x\right]\left(1-\frac{s^2}{c^2}x^2\right)^2,\\
 B_2(x;n)=&\frac{n+2}{n}\frac{s^2}{c^2}x^2\left[1+\left\{|\alpha|^2-|\beta|^2-\frac{(\alpha\overline{\beta}+\overline{\alpha}\beta)s(3s^2+1)}{c^3}\right\}x\right.\nonumber\\
 &\left.-4\left\{|\alpha|^2-|\beta|^2+\frac{(\alpha\overline{\beta}+\overline{\alpha}\beta)s}{c}\right\}\frac{s^2}{c^2}x^3+\frac{4(\alpha\overline{\beta}+\overline{\alpha}\beta)s^3}{c^3}\frac{x}{1-x^2}\right],\\
 B_3(x;n)=&\frac{n+2}{n-2}\left[1+\left\{|\alpha|^2-|\beta|^2+\frac{(\alpha\overline{\beta}+\overline{\alpha}\beta)s}{c}\right\}x\right]\frac{s^2}{c^2}x^2\left(1-\frac{s^2}{c^2}x^2\right),
\end{align}
and $I_A(x)=1$ if $x\in A$, $I_A(x)=0$ if $x\notin A$.
By Eq. (\ref{eq:r-th_mom}), we can compute the characteristic function $E(e^{izX_t/(n+2)\tau})$ as $\tau\rightarrow\infty$.
Thus the calculation of the density function for $m=2,n=3,4,\ldots$ case is completed.
\begin{flushright}
$\Box$ 
\end{flushright}

\begin{theacknowledgments}
The author is grateful to Joe Yuichiro Wakano and the Meiji University Global COE Program ``Formation and Development of Mathematical Sciences Based on Modeling and Analysis'' for the support.
\end{theacknowledgments}

\clearpage

%%%%%%%%%%%%%%%%%%%%BIBTEX%%%%%%%%%%%%%%%%%%%%%%%%%%%%%%%



\bibliographystyle{aipproc}   % if natbib is available

%\bibliography{main}

\IfFileExists{\jobname.bbl}{}
 {\typeout{}
  \typeout{******************************************}
  \typeout{** Please run "bibtex \jobname" to optain}
  \typeout{** the bibliography and then re-run LaTeX}
  \typeout{** twice to fix the references!}
  \typeout{******************************************}
  \typeout{}
 }

\end{document}